\documentclass[]{aa}  

\usepackage{graphicx}
\usepackage{color}
\usepackage{ragged2e}
\usepackage{txfonts}
\usepackage{hyperref}
\usepackage{euclid}
\usepackage{orcidlink}
\usepackage[switch, modulo]{lineno}
\newcommand\rev[1]{\textcolor{black}{ #1}} 
\newcommand{\orcid}[1]{\orcidlink{#1}}
\begin{document}

   \title{Multiscale exploration of SMACS J0723.3--7327's intracluster light and past dynamical history }

   \author{Amaël Ellien\orcid{0000-0002-1038-3370}\inst{1}}

   \institute{OCA, P.H.C Boulevard de l'Observatoire CS 34229, 06304 Nice Cedex 4, France}

   \date{Submitted 09/01/2025; Accepted for publication in A\&A}

 
  \abstract
{In this work an analysis of the intracluster light (ICL) in the galaxy cluster SMACS~J0723.3–7327 (hereafter, SMACS~J0723) using JWST/NIRCam deep imaging in six filters (F090W to F444W) is presented. The images were processed for low surface brightness (LSB) science, with additional correction for instrumental scattering in the short-wavelength channels, and analysed using wavelet-based decomposition. The ICL, brightest cluster galaxy (BCG), and satellite galaxies were extracted and modelled, with 2D maps for each component.
ICL and ICL+BCG fractions, computed across all filters within a 400\,kpc radius, exhibit a flat trend with wavelength, averaging 28\% and 34\%, respectively. Flux ratios between the BCG and the next brightest members (M$_{12}$, M$_{13}$ and M$_{14}$) also display minimal wavelength dependence. These results indicate that SMACS~J0723 is a dynamically evolved cluster with a dominant BCG and well-developed ICL.
Five prominent ICL substructures are analysed, contributing to 10–12\% of the total ICL+BCG flux budget, slightly exceeding simulation predictions. Their short dynamical timescales suggest an instantaneous ICL injection rate of several $10^3 L_{\odot}\,{\rm yr}^{-1}$, consistent with active dynamical assembly. These findings support a scenario where SMACS~J0723's ICL growth is currently driven by galaxy mergers involving the BCG and other bright satellites, rather than by the accretion of pre-processed ICL from a recent cluster merger. However, extrapolating the current injection rate to the cluster's lifetime indicates that additional mechanisms are required to match the growth observed in other clusters over cosmic times.}


\maketitle

\begin{figure*}
    \centering
    \includegraphics[width=\textwidth]{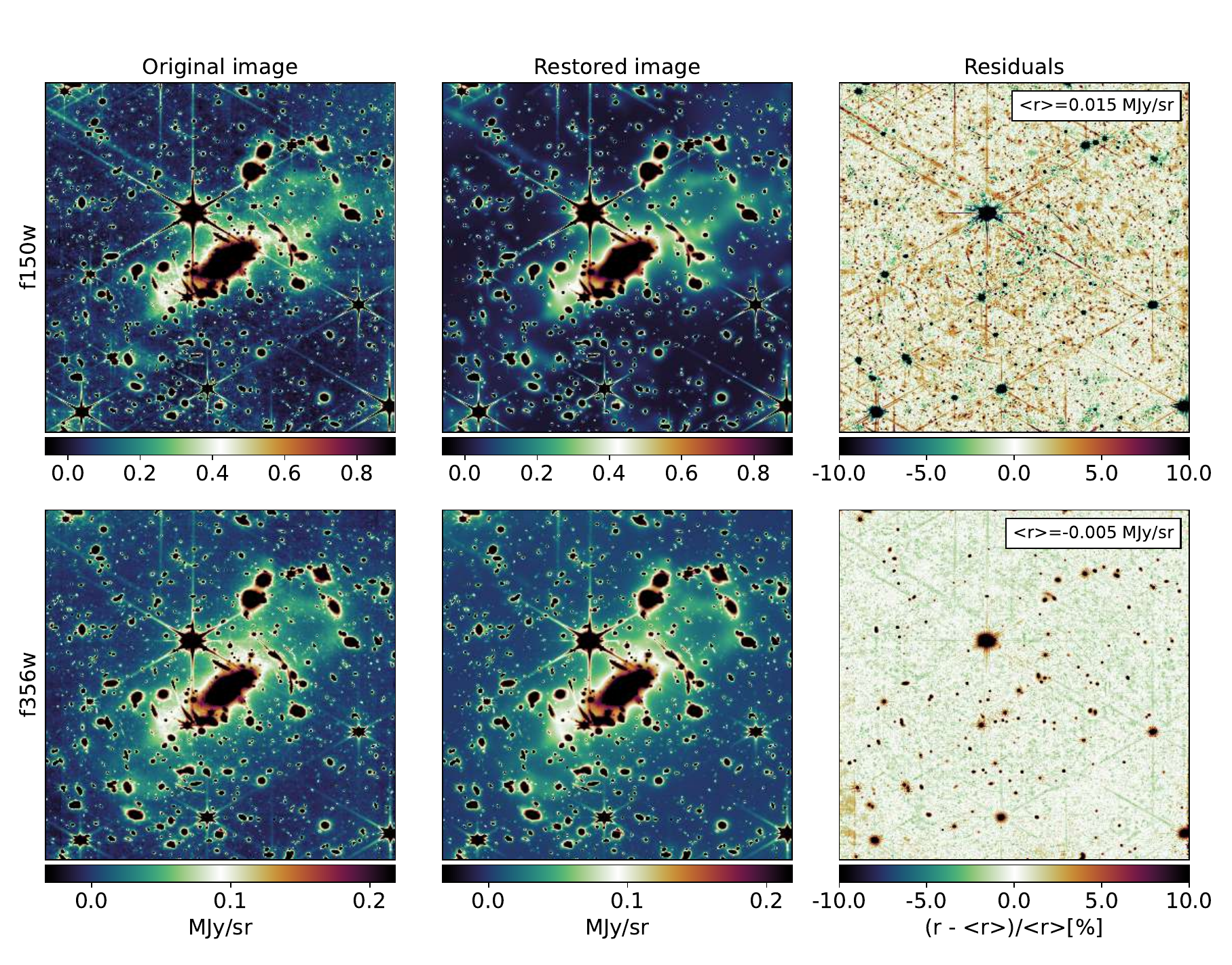}
    \caption{\texttt{DAWIS} outputs for the F150W and F356W filters, shown respectively on the \rev{top and bottom panels}. From left to right: original field, complete restored field by \texttt{DAWIS} and residuals.
    \rev{The size of the images is $\ang{;1.75;}\times\ang{;1.75;}$ ($\sim680\,\mathrm{kpc}\times680\,\mathrm{kpc}$ at the cluster's redshift).}
    The residuals are displayed with enhanced contrast (\rev{deviation of individual residual pixels $r$ from the median residual value $<r>$, expressed in percent)}, demonstrating the excellent reconstruction quality of the wavelet process over the whole image.}
    \label{fig:plot_array_mix}
\end{figure*}
\begin{figure*}
    \centering
    \includegraphics[width=\textwidth]{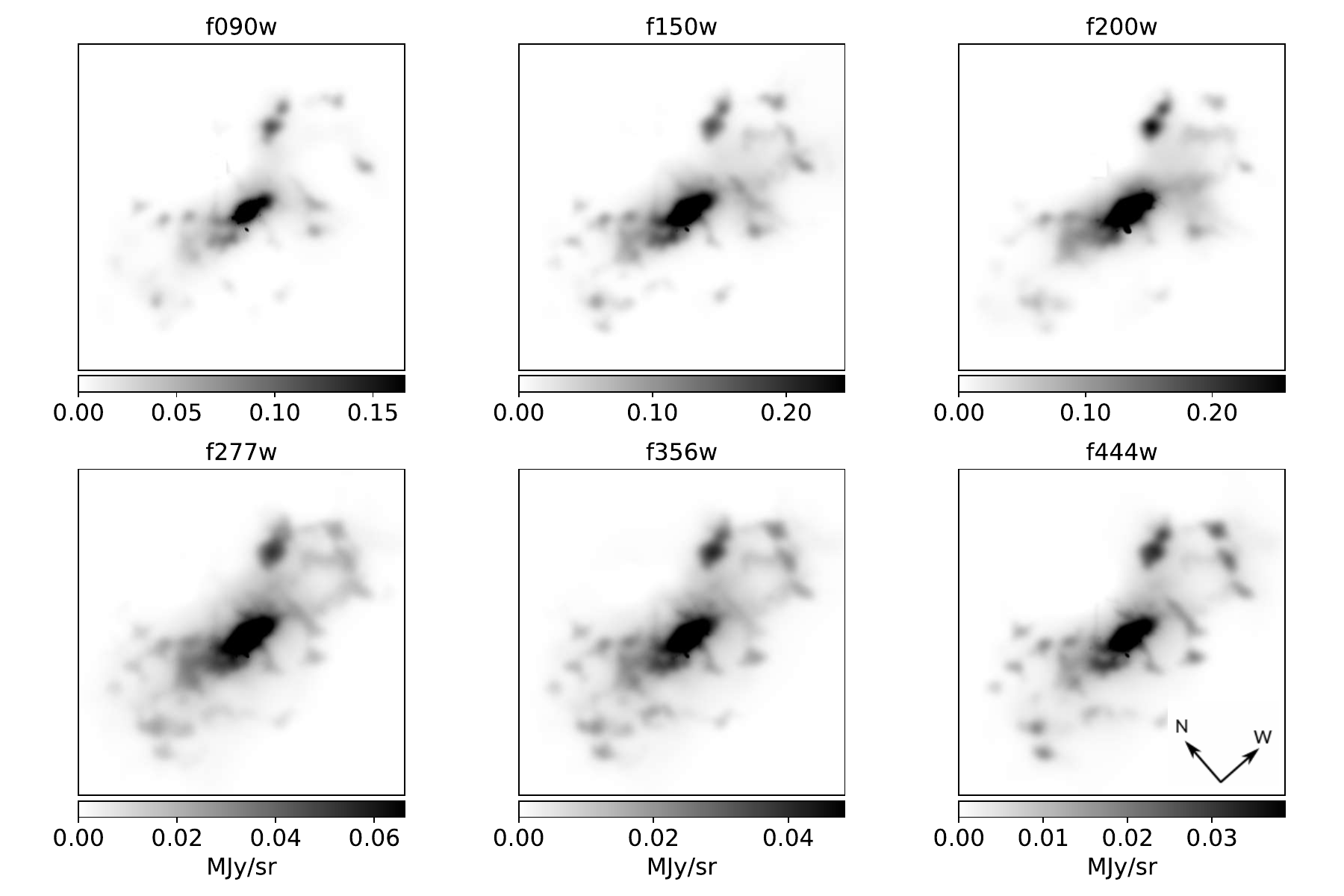}
    \caption{BCG+ICL synthesis maps for each filter resulting from a combined \rev{wavelet separation (WS) + size separation (SS) + spatial filtering (SF)} scheme with $w_{\mathrm{s}}=5$ and $S_{\mathrm{a}}=80$~kpc. \rev{The size of the images is $\ang{;1.75;}\times\ang{;1.75;}$ ($\sim680\,\mathrm{kpc}\times680\,\mathrm{kpc}$ at the cluster's redshift).
    To highlight the diffuse emission, the scale for each map is made from the corresponding original image with the \textit{zscale} algorithm from \texttt{DS9}\citep{Joye2003}.
    The ICL+BCG distribution looks similar for all filters except the F090W, where substructures are missing.}}
    \label{fig:plot_example_recim_all_filter_maps}
\end{figure*}

\section{Introduction}

In the current $\Lambda$CDM paradigm, galaxy clusters assemble hierarchically through the accretion of other clusters and groups. 
During this non-linear process, the galaxies evolving in these dense environments experience many gravitational interactions, with other cluster galaxies or with the cluster potential itself. 
One of the outcomes of these dynamical interplays is the intracluster light (ICL), a faint \citep[$\mu_V\geq26.5$\,mag.arcsec$^{-1}$,][]{Rudick2006} feature of galaxy clusters emitted by stars bound to the cluster potential rather than individual galaxies \rev{\cite[see][for reviews]{Montes2022revue, Contini2022}. }
Observations of nearby, well resolved, galaxy clusters such as Virgo and Coma best reveal the ICL for what it is: a smooth and diffuse halo \cite{Jimenez2019} on top of substructures such as tidal streams \cite{Mihos2005, Janowiecki2010, Mihos2017}, plumes \cite{Gregg1998} or curved arcs \cite{Trentham1998}.
Each of these tidal debris is a fossil of a past dynamical interaction, encoding in its properties (shape, morphology, colour or again spatial position within the cluster) information about the time the event occurred or the actors involved in it. 
Dissecting the ICL and its subcomponents is therefore akin to conducting an archaeological investigation of a cluster's dynamical history.

Numerical studies early pointed to tidal stripping of stars by the gravitational potential of the cluster \cite{Byrd1990} and violent galaxy-galaxy interactions \citep{Moore1996, Moore1999} to produce tidal streams in galaxy clusters.
Following works using $N$-body simulations acted multiple steps during the formation of ICL, with the creation of transient, dynamically cold streams, followed by their dissolution into the cluster potential \citep{Rudick2006, rudick2009}.
This formation channel, denoted as stellar stripping, has been shown as an efficient way to inject stars into the ICL \citep{Contini2014, Contini2018}, although the exact physical processes (major/minor galaxy merger, fly-by interactions..) and the link to the resulting tidal feature properties are not understood in detail.
Growing computational power has allowed simulations to probe hundreds to thousands of galaxy clusters \citep{Remus2017, Springel2018, Pillepich2018, Nelson2024}, demonstrating the stochastic nature of ICL formation \citep{Contini2023} and the dominant contribution of stellar stripping relative to other formation channels \citep{Chun2023, Chun2024, Contini2024}. However, pre-processing \citep{mihos2004} contribution is also significant.
It is relevant here to note the recent works of \cite{Martin2024}, who highlighted the importance of refined mass resolution in $N$-body simulations, as poorly resolved dark matter halos could lead to excessive stellar stripping.


On the observation side, the overall ICL fraction, morphology and colour have been studied in galaxy clusters, from $z\sim0.1-0.5$ \citep[non-exhaustive list:][]{Jimenez-Teja2018, Montes2019, Kluge2020, Ellien2019, Zhang2019, Furnell2021, Dupke2022, Jimenez-Teja2023, Ragusa2023, golden-marx2024, kluge2024} up to $z\sim1-2$ \citep{adami2013, Joo2023}. 
Despite the pervasive nature of the ICL, the importance of the different formation mechanisms, the growth rate of ICL with redshift, and its dependence on cluster properties are still discussed.
Tidal substructures are almost absent from this literature because they blend in the diffuse ICL for clusters at $z \ge 0.1$ due to cosmological dimming and decreased spatial resolution.
The recent advances in near-infrared (NIR) instrumentation, particularly with the launch of the \textit{James Webb Space Telescope} (JWST) provide an unprecedented opportunity to fill this gap and improve our knowledge of ICL tidal debris at \rev{$z\geq0.2$} and their link to the dynamical history of their host cluster.
This opportunity arose promptly with the JWST Early Release Observations \citep[EROs; ][]{Pontoppidan2022} of SMACS~J0723, a cluster with a mass \rev{$M_{500}=7.9\pm1.1\times10^{14}M_{\odot}$ \citep{Finner2023}}
, at $z=0.3877$, part of the MACS sample \citep{Repp2018}.
This cluster displays \rev{a brightest cluster galaxy (BCG) at position \mbox{${\rm RA}=\ang{110.82734;;}$}, \mbox{${\rm Dec}=\ang{-73.454683;;}$,}} and large amounts of ICL up to 400\,kpc in JWST images, with several visible tidal features \citep{Montes2022jwst}. 
By combining multi-wavelength data, \citet{Mahler2023} provided evidence that this cluster may have experienced a recent merging episode, probably on an axis close to our line of sight.
Subsequent works showed that while SMACS~J0723's ICL correlates with globular cluster and dwarf galaxy density distributions \citep{Martis2024}, fine substructures and cavities in the ICL were not found in the dark matter-dominated mass distribution \citep{Diego2023}.

This work focuses on constraining the past and current dynamical state of SMACS~J0723 using different markers: ICL fractions, magnitude gap and tidal stream properties.
The open-access data used for the analysis are briefly described in Section~\ref{sec:data}. Section~\ref{sec:image_analysis} contains the description of the methods used to model the different components of the images, and the results are shown in Section~\ref{sec:results}. A summary of the paper is provided in Section~\ref{sec:conclusions}. A standard $\Lambda$CDM cosmology is assumed, with $\Omega_{m}=0.3$, $\Omega_{\Lambda}=0.7$ and $H_{0}=70$~km\,s$^{-1}$\,Mpc$^{-1}$. This results in a scale of $5.271$~kpc/$\arcsec$.

\section{Data}
\label{sec:data}

\subsection{NIRCam images}
\label{subsec:images}

MACS~J0723 has been observed with the NIRCam in six different filters: short channel (F090W, F150W and F200W) and long channel (F277W, F356W, and F444W). While the exquisite sensitivity of the JWST ensures deep photometric limits, the released reduced images were unsuitable for low surface brightness (LSB) astronomy due to instrumental light gradients.
\rev{For long channel images, \citet{Montes2022jwst} corrected background gradients in individual frames by masking all galaxies, foreground and background objects, and the cluster's core. They then fitted a second-degree 2D polynomial to the masked frames to model and remove the gradient while preserving the ICL. After refining the background correction through an iterative masking process, they co-added the corrected frames and subtracted a constant background value from the final images.}
\rev{The short channel images are trickier since the detectors are smaller than the cluster’s ICL extent and lack a common gradient, the authors avoided fitting a plane and instead corrected for readout patterns and detector scaling. They first removed low-level electronic noise (`1/f noise') by subtracting each row's $3\sigma$-clipped median value, then scaled the four detectors using a reference detector before assembling the final mosaics.}
The images used in this work are the corrected images, available in open access\footnote{\url{https://www.dropbox.com/sh/zzb7fl1j0vl1vit/AABW1uz06Vn2XGCYAmOJIbega?dl=0}}. Despite the improved pipeline, the short channel images \rev{still present 1/f noise residuals as stripe patterns}, and \citet{Montes2022jwst} did not use them in their analysis.
\rev{In this current work, additional attention is given to further correct for this instrumental background (see Sect.~\ref{subsec:preprocessing} and Appendix~\ref{app:short_channel_scattered_light_background})}.

Because these data were part of the original release \citep{Pontoppidan2022}, photometric calibration systematics were issued for this version of the images \citep{rigby2023}.\rev{
As a result, absolute flux values cannot be reliably compared across different filters. 
To mitigate this limitation, all flux-involved quantities computed and analysed in this work are expressed in dimensionless forms, such as ICL fractions and flux ratios, ensuring consistency despite calibration uncertainties.}

\subsection{Galaxy catalogue}
\label{subsec:galaxy_catalogs}

Several galaxy catalogues of the MACS~J0723 field are already available in the literature. Most notably the cluster's red sequence (RS) has been identified by \citet{Mahler2023} using Hubble Space Telescope (HST) F606W and F814W filters, resulting in a catalogue of 144 galaxies. They also independently identified 26 galaxies with spectroscopic redshifts in VLT/MUSE data, all of which but four belong to the RS. 
More recently, \citet{Noirot2023} published a spectroscopic redshift catalogue of 190 sources, measured using joined MUSE, NIRISS and NIRSpec data. 
For the present work, galaxy members are extracted from the \citet{Noirot2023} catalogue by keeping sources with redshift within $0.3877\pm0.01$. 
This results in a catalogue of 45 galaxies, of which 38 are already part of the RS identified by \citet{Mahler2023}. 
The final galaxy member catalogue is created by concatenating the 144 galaxies of the RS, the 4 VLT/MUSE galaxies identified by \citet{Mahler2023}, and the 7 new members identified by \citet{Noirot2023}. The final galaxy member catalogue contains 155 sources. 


\section{Image analysis}
\label{sec:image_analysis}

A look at the NIRCam images of MACS~J0723 is enough to grasp the difficulties associated with ICL analysis. Not only the density of sources is extremely high, resulting in many overlapping light halos, but there is also a great variety of shapes and morphologies (foreground MW star halos and crosses, strongly lensed background galaxies...). The ICL does not have a smooth distribution but is rather highly structured with peculiar features such as several tidal streams. The prior-free\footnote{In the sense that no particular shape is assumed for sources.} multiscale decomposition provided by the wavelet-based method DAWIS \citep[standing for Detection Algorithm with Wavelets for Intracluster light Studies;][hereafter E21]{Ellien2021} is therefore particularly \rev{well-suited} to model such complex astronomical fields and decouple the ICL distribution from other light sources.


\subsection{Preprocessing}
\label{subsec:preprocessing}

Before applying \texttt{DAWIS} to the images, a few processing steps are performed to facilitate the wavelet analysis. Using the \texttt{GNU Astronomy Utilities} \texttt{astwarp} routine \citep{Akhlagi2015, Ashofteh-Ardakani2023} the images are rotated by $35\deg$, so the axes of the image correspond to the two directions of the 2D wavelet transforms. For consistency and to diminish computing time, short-channel images are also resampled to the same pixel size as long-channel images. The astronomical field is then cropped with the \texttt{astcrop} routine from a $2215\times2215~\mathrm{pix}^2$ into a square image of size $2048\times2048~\mathrm{pix}^2$ \rev{(corresponding to size $\sim680\,\mathrm{kpc}\times680\,\mathrm{kpc}$ at the cluster's redshift)}. This size conveniently corresponds to the wavelet scale $n=11$, and removing the outskirts minimizes effects due to visible differences in noise statistics close to the image's original border. Finally, the instrumental light background of short channel images is modelled as a combination of a linear sharp-edge component and a large-scale background. This combined background is removed from the images, before the wavelet analysis presented in the next sections. 
After correction, the limiting depth \rev{($3\,\sigma$, $10\,\arcsec\times10\,\arcsec$)}\footnote{Computed in the same way as \citet{Montes2022jwst}.} of the short channel images jumps down to values similar to the long channel images ($\sim31$~mag\,arcsec$^{-2}$), which is an improvement over the limiting depth before the correction ($\sim28$ to $29.5$~mag\,arcsec$^{-2}$). A complete description of the methodology used for the instrumental light background and the corresponding images can be found in Appendix~\ref{app:short_channel_scattered_light_background}.
 
\subsection{Multiscale decomposition}
\label{subsec:DAWIS}

Most of the analysis is performed with \texttt{DAWIS}, an automated wavelet-based detection algorithm optimized for LSB astronomy. The algorithm is already extensively described in \citet{Ellien2021}, so only salient points are summarized here. \texttt{DAWIS} decomposes all the image information content into light contributions denoted ``atoms''. The process is performed iteratively, each iteration following the steps of a multiresolution vision model \citep{Bijaoui1995}: i) a wavelet transform is applied to the analysed image, ii) regions of significant wavelet coefficients are detected and identified as sources, iii) the light profiles of the brightest sources (set by a relative threshold $\tau$) are restored from their wavelet coefficients. An attenuation factor $\gamma$ is then applied to each restored light profile to diminish potential restoration mistakes, and the resulting restored 2D light profile is the so-called atom. The atoms are removed from the image, and the algorithm reiterates the whole process on the residuals upon convergence. The algorithm convergence is controlled by a parameter $\epsilon$ (see Appendix~\ref{app:dawis_input_parameters} for details about the parameters and their values for this work). 

The output of \texttt{DAWIS} for each image is a list of atoms, the sum of which reproduces the entire astronomical field. The resulting restored image and residuals are displayed for the F150W and F356W filters in Figure~\ref{fig:plot_array_mix}. 
Due to the nature of the isotropic B3-spline wavelet used by \texttt{DAWIS}, some bright and/or linear features are not captured with as much quality as others.
\rev{Typically, bright foreground stars (including their cores and diffraction spikes) are excessively restored by the algorithm, leading to overestimating their flux. 
This occurs because the algorithm injects too much flux into the individual elements composing these sources, indicating that additional iterations are required to model their complex structures. Similarly, the cores of some bright galaxies appear overly spread due to this effect.}
However, the overall restoration quality is excellent with values under $1\%$ of the average in the residuals (equivalent to differences of the order of $10^{-5}$~MJy/sr between the original image and the synthesis image).

\subsection{Atom selection and image synthesis}
\label{subsec:atom_selection}

One atom does not represent an entire astrophysical source, but rather a substructure. Note that such a decomposition is artificial and depends on how the signal was captured in the wavelet space at each iteration. However, atoms can be selected to synthesize images containing the light profile of specific astrophysical sources, the most trivial case being the sum of all atoms detected and restored by \texttt{DAWIS}, which results in a synthesis image of the entire astronomical field (see previous section). 

In this work several atom properties are used as priors depending on the selection: the wavelet scale $w_{\mathrm{s}}$ at which it has been detected (wavelet separation, WS), the coordinates of the local maximum of the atom (spatial filtering, SF) and the size $S_{\mathrm{a}}$ of the atom (size separation, SS). A large range of selection schemes and parameter values are applied to produce the following synthesis images for each filter: ICL, ICL+BCG, member galaxy (with and without the BCG), and total cluster (ICL+BCG+satellites) maps. The details about the different selection schemes, the effects of their parameters, and the synthesis of ICL maps are given in Appendix~\ref{app:atom_selection_schemes}.

\section{Results}
\label{sec:results}

Several ICL and ICL+BCG maps are produced using different atom selection schemes. For simplicity, the combined WS+SF+SS scheme with $w_{\mathrm{s}}=5$ and $S_{\mathrm{a}}=80$~kpc is kept as the main analysis scheme for this work and is discussed here. This choice is detailed in Appendix~\ref{app:atom_selection_schemes}. The corresponding ICL+BCG  maps are shown for each filter in Figure~\ref{fig:plot_example_recim_all_filter_maps}. All discussed values here are computed from these maps.

\subsection{ICL 2D distribution and fractions}

\label{subsec:icl_dsitributions_fractions}
A visual inspection of the models \rev{shown in Fig.~\ref{fig:plot_example_recim_all_filter_maps}} shows that the complex multiscale nature of the ICL, featuring an extended 2D distribution over hundreds of kpc spanned by substructures, is well captured by the \texttt{DAWIS} wavelet modelling. 
The ICL distribution appears similar in all filters except F090W, where the faint diffuse signal is not as well recovered. 
\rev{Indeed, the F090W ICL+BCG map is missing substructures compared to the other bands, which is expected considering this image had the lowest photometric depth while being the most impacted by instrumental light contamination}.
\rev{In each filter, a bright compact source to the south-east of the BCG can be seen. This signal is neither ICL nor BCG, and shows that some non-ICL, non-BCG atoms can be accidentally included in the model.
For this reason, a bootstrap procedure on atom selection is applied when making any measurements on the synthesized maps (see Appendix~\ref{app:atom_selection_error}).}

The average diffuse ICL distribution is elliptical, with a major axis in the east-west direction. The BCG and ICL major axes appear well aligned with a positional angle in the range $\ang{35;;}$-- $\ang{41;;}$ through all filters. The ICL ellipticity is higher than that of the BCG in every filter ($\sim0.26$--$0.43$ against $\sim0.41$--$0.57$ ). \rev{This supports the idea that while the ICL and BCG are different components \citep{Contini2018}, they are correlated and co-evolve as the transition from the BCG profile to the ICL is continuous.} Although the SS criterion applied ($S_{\mathrm{a}}=80$~kpc) is consistent with typical values found in the literature to mark this transition \citep{Contini2022, Brough2024}, it is artificial and probably not separating two physical components. This is also supported by the rather high ICL fractions relative to the ICL+BCG fractions (see next paragraph).

\begin{figure}[h]
    \centering
    \includegraphics[width=\columnwidth]{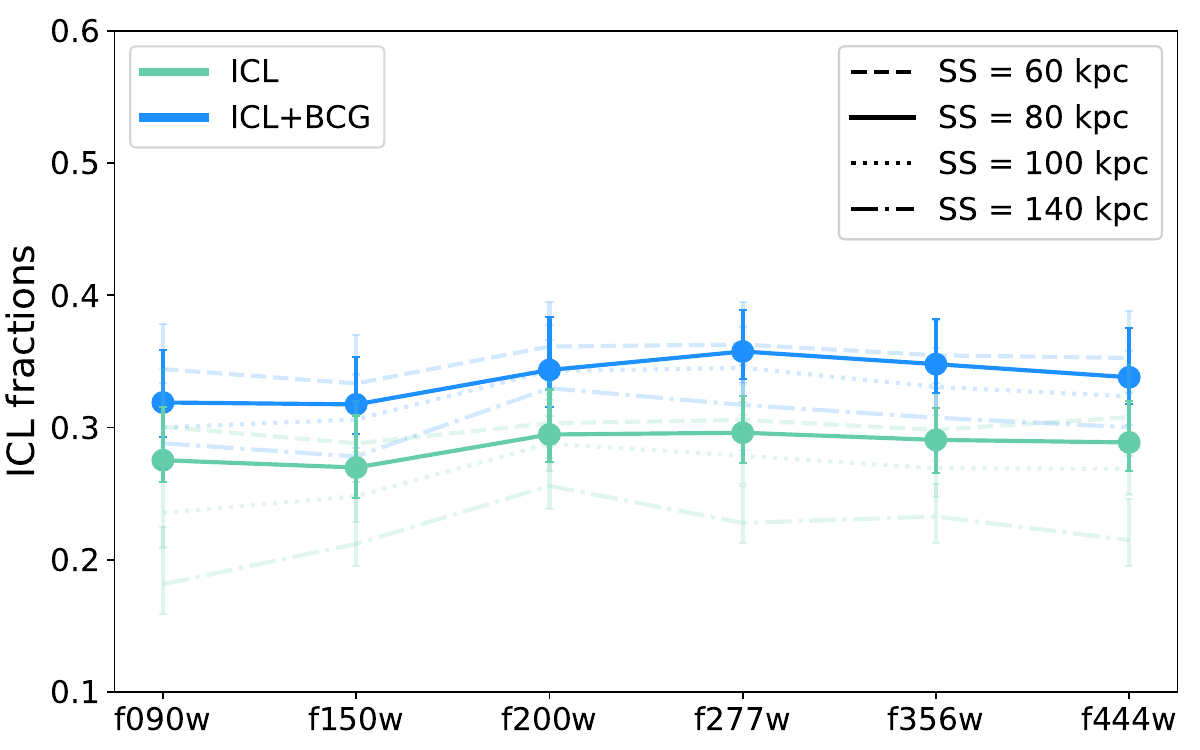}
    \caption{ICL and ICL+BCG fractions for each filter resulting from a combined WS+SF+SS scheme with $w_{\mathrm{s}}=5$ and $S_{\mathrm{a}}=80$~kpc.}
    \label{fig:fICL_vs_filters}
\end{figure}

The ICL (and ICL+BCG) fractions and their error bars are computed for each filter and each selection scheme with the bootstrap procedure described in Appendix~\ref{app:atom_selection_error}. The resulting fractions for a combined WS+SF+SS scheme with $w_{\mathrm{s}}=5$ and $S_{\mathrm{a}}=80$~kpc are displayed in Figure~\ref{fig:fICL_vs_filters}. Both the ICL (ICL+BCG) fractions are almost completely flat within error bars over the six NIRCam filters, with an average value of $28\%$ (respectively $34\%$). The F090W and F150W fractions seem slightly lower than others, which can be explained by an over-subtraction of the instrumental light background (see Appendix~\ref{app:short_channel_scattered_light_background}), but the differences are not significant considering the error bars. These values are consistent with values found in the visible/NIR bands in the literature for clusters in the $0.1$--$0.5$ redshift range, although on the upper end of the distribution \citep[see Figure~2 of][]{Montes2022jwst}.
\rev{As different stellar populations would produce different spectral energy distributions, and therefore a variation of the ICL fraction with wavelength, the flat fractions here indicate an intracluster stellar population with similar properties to satellite galaxy stars. 
This suggests galaxy stripping is the main ICL formation mechanism on the cluster scale.
This is true independently of the selection scheme used to produce the map.}

\subsection{Tidal features}
\label{subsec:tidal_features}

\begin{figure}[h]
    \centering
    \includegraphics[width=\columnwidth]{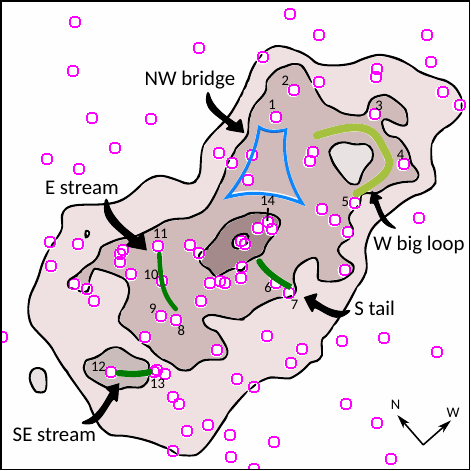}
    \caption{Schematic view of MACS~J0723's stellar content and substructures. The filled contours are drawn from the F356W ICL+BCG map from DAWIS and purple circles are the satellite galaxies (see Sect.~\ref{subsec:galaxy_catalogs}). Galaxies discussed in the main text are indicated by a number (1--\rev{14}, see Sect.~\ref{subsec:tidal_features}). The main tidal features are highlighted with coloured annotations: the east, south-east and south tidal streams are shown in dark green, the external section of the West big loop in lime green, and the north-west bridge in blue.}
    \label{fig:plot_substructures_wavelets}
\end{figure}

A common step in  ICL formation through gravitational interactions between galaxies (satellite and/or the BCG) is the creation followed by the dissolution of tidal features \citep{rudick2009}. 
These are seen in nearby galaxy clusters such as Virgo \citep{Mihos2005, Mihos2017} or Coma \citep{Trentham1998, Gregg1998}, but are usually more difficult to see at higher redshifts, due to cosmological dimming and to the lower spatial resolution.
Thanks to JWST's great sensitivity, MACS~J0723 provides a great opportunity to study such tidal features at intermediate redshift, as its ICL is extensively clumpy with prominent over-densities.
This analysis focuses on five diffuse substructures with different morphologies and sizes. They are detected in every filter (except F090W, where the faint diffuse signal is not recovered as well as in the other bands), implying the stellar populations in the tidal features are not very different from the average intracluster and satellite galaxy stellar populations and originate from galaxy stripping. The positions of these structures are shown schematically in Fig.~\ref{fig:plot_substructures_wavelets} superposed on the contours of the ICL+BCG map in the F356W filter and satellite galaxy positions. Most of these features are already denoted by \citet{Montes2022jwst}, namely: the big west (W) loop, the large north-west (NW) bridge, and the east (E) and south-east (SE) tidal streams. Here, an additional tidal tail in the South (S) of the BCG is added to the list and is denoted as the S tail.

The largest feature is the NW bridge, with an apparent size $\sim 75 {\rm\,kpc}\times 50 {\rm\, kpc}$. It links the BCG (and the other interacting satellite in its close halo) with galaxies 1 and 2, respectively the second and third brightest galaxies in the cluster. This indicates that these two massive elliptical galaxies are undergoing strong tidal stripping by the BCG, with a recent close encounter contributing significantly to the build-up of the ICL. The fraction of flux enclosed in the NW bridge to the total ICL flux budget yields values of 3--5\% through each filter.

The second largest ICL substructure is the W big loop, with a focus here on its outer arc, which has a length of $\sim150\,{\rm kpc}$ from end to end and a width of $\sim30 {\rm\, kpc}$. This unique feature (to the current author's knowledge, no other loop of this sort has been detected in the ICL of another galaxy cluster) is well aligned with the major axis of the total ICL distribution. It contributes to 2--4\% of its total flux budget (except for the F090W filter, where it is not detected and has a budget close to null). It has been identified as a single stellar stream by \citet{Montes2022jwst}, due to its flat colour over kpc scales.
However, the origin of the big loop is puzzling as it does not look tied to any satellite galaxy; the two large galaxies it seems to be associated with \rev{have very different colours \citep{Montes2022jwst}, and} are foreground sources that do not belong to the cluster. The satellite galaxies 3 and 4 could be spatially associated with the large loop, but are smaller (diameter $<10\,{\rm kpc}$) than the width of the big loop itself, excluding them as the main origin for this very large tidal structure.
\citet{Diego2023} pointed out that while the cavity is not seen in the total mass 2D distribution, this difference could be attributed to the different nature of the baryonic and dark matter components, and result from strong stripping of satellite galaxies following close encounters with the BCG.
The dynamical timescale inferred by the mass enclosed in a 200\,kpc radius \citep[$M_{200\,{\rm kpc}}\sim 1.5\times 10^{14} M_{\odot}$,][]{Mahler2023} is $\sim 0.2$\,Gyr. This provides a characteristic decay timescale for tidal streams of $\sim 0.3$\,Gyr \citep{rudick2009} at these distances from the cluster centre. Assuming the average distance between the cluster centre and the big loop is $200$\,kpc (a reasonable assumption although this radius can vary by a few tens of kpc, depending on what part of the loop is considered), this suggests that it must have formed recently ($<0.3$\,Gyr).
For the BCG to be at the origin of the W big loop, a collision differential velocity of $\sim1000$\,km\,s$^{-1}$ with the other satellite progenitor is needed, assuming the loop conserved a similar speed after being created.
This is well within the range of characteristic velocity dispersion of galaxies in clusters \citep[][]{Goto2005}.
Assuming the collision happened perpendicular to the line of sight and that minimal projection effects are at play (a strong hypothesis), the most probable other progenitor is \rev{galaxy 14,} the elliptical galaxy close to the BCG and currently merging with it.

The E and SE streams have apparent sizes of $\sim 75 {\rm\, kpc}\times 15 {\rm\, kpc}$ and $\sim 40 {\rm\, kpc}\times 30 {\rm\, kpc}$ respectively. Overlooking projection effects, the SE stream seems to result from a past encounter between galaxy 12 and galaxy 13. Similarly, the E stream appears to be associated with galaxies 8--11. They do not contribute as much as the W big loop or the NW bridge to the total ICL flux budget ($\sim$1--2\% each), but are textbook cases of stellar stripping through violent encounters between satellite galaxies.
Lastly, the S tail also has a rather small size ($\sim 30 {\rm\, kpc}\times 10 {\rm\, kpc}$), and appears to originate from the interaction of galaxies 6 and/or 7 with the BCG. Its morphology appears more clumpy than the E and SE streams with patches of stars, indicating a more recent interaction and a less advanced decay state (although it could decay faster due to its proximity to the cluster centre). The light fraction in the S tail to the total ICL is similar to the streams, with values around $\sim$1\%.

\subsection{Magnitude gap}

Another way of probing the dynamical state and age of the cluster is through the prism of the BCG properties contrasted with the other satellite galaxy members. As the BCG is expected to continuously accrete satellites through its lifetime  \citep{Solanes2016}, the magnitude differences between the BCG and the second (hereafter M$_{\rm 12}$), third (M$_{\rm 13}$) or fourth (M$_{\rm 14}$) brightest cluster members are hints on the advancement of this build-up \citep{Milosavljevic2006}. A large magnitude gap indicates an old, well-evolved system, with a dominant BCG; the opposite traces perturbed and merging systems.

The BCG and second to fourth satellite magnitudes are estimated in each band by integrating the flux from the corresponding \texttt{DAWIS} output models. As stated in Sect.~\ref{subsec:icl_dsitributions_fractions}, the performed separation between BCG and ICL probably underestimates the BCG component. All values discussed here are therefore lower limits on the magnitude gap. For each filter, the measured values for the second and third brightest satellites are found in the range M$_{\rm 12}\sim 1.2-1.3$\,\rev{mag} and M$_{\rm 13}\sim1.3-1.6$\,\rev{mag}. These are galaxies 1 and 2 as indicated in Fig.~\ref{fig:plot_substructures_wavelets}, linked to the BCG by the NW bridge. As discussed in the previous section, these galaxies have probably undergone a past encounter with the BCG, and will probably end up merging with it. The fourth brightest member gap is higher with M$_{\rm 14}\sim 2.3-2.5$\,\rev{mag} in all filters. The corresponding galaxy is the elliptical galaxy close to the BCG centre and merging with it, as noted by \citet{Montes2022jwst}.

Considering they are lower limits, these values indicate a dominant and well-evolved BCG, positioned at the bottom of the cluster potential well, and currently accreting a large quantity of matter from its satellites. This is consistent with the high ICL and ICL+BCG fractions measured in Sect.~\ref{subsec:icl_dsitributions_fractions}, as both are linked to the BCG hierarchical assembly and therefore correlate \citep{golden-marx2024}.

\section{Discussion}

\subsection{Multiple steps in the ICL formation}

It has now been accepted that the main formation channels of the ICL at $z<1$ are pre-processing and galaxy stripping \citep{Contini2021}, with different importance depending on the cluster's current dynamical state.
These mechanisms leave different imprints on the colour of the ICL. A negative radial colour profile is indicative of an intracluster stellar population following the galaxy morphological segregation, and is therefore associated with the stripping of the outskirts of Milky Way-like galaxies infalling in the cluster (this is the case for MACS~J0723 at $r>100$\,kpc, as identified by \citet{Montes2022jwst}). On the other hand, pre-processing results in local colour variations due to inhomogeneities in the ICL right after a cluster major merger even though it is hard to track once the system has had time to relax. Ongoing stripping of galaxies is hard to catch in images due to the transient nature of tidal streams in galaxy clusters, their LSB nature and their low contrast on the diffuse ICL. This dataset is therefore a great opportunity to directly investigate the intermediate steps in forming the ICL of SMACS~J0723 through the properties of the tidal features in its ICL.

Altogether, the tidal features account for approximately 10--12\% of the total ICL+BCG budget ($\sim$8\% for the F090W without the big loop) within a 400\,kpc radius. As simulations predict on average 5 to 10\% of the ICL material in stellar streams \citep{rudick2009}, SMACS~J0723 appears to have a high fraction of its ICL in non-relaxed tidal features. 
All \rev{SMACS~J0723's ICL} substructures have crossing and decay timescales lower than 0.5\,Gyr. 
\rev{The associated star injection rate (the rate at which stars are added to the ICL through dynamical processes) is computed by taking the total flux enclosed in each tidal structure and dividing it by the upper limit of the crossing time (0.5\,Gyr).
This provides an averaged star injection rate of the order of a few $10^3 L_{\odot}.{\rm yr}^{-1}$ in every filter over this period.}
This is a lower limit, but it is still three orders of magnitude higher than values found in the nearby Virgo cluster \citep[$\sim 1\, L_{\odot}\,{\rm yr}^{-1}$,][]{Mihos2017}, supporting the fact that SMACS~J0723 is in the process of building-up its ICL+BCG system, with an undergoing phase of strong growth. Assuming this injection rate to be constant through the life of the cluster yields a lower limit growth \rev{factor} 
of $\sim 1.7$ between $0.1<z<0.5$. This is slightly lower than what can be seen in other clusters, that have an average ICL growth \rev{factor} of $\sim$2--4 \citep{Furnell2021}. This is expected considering it does not include all ICL injection mechanisms contributing to the growth of the ICL during a cluster lifetime (pre-processing, mergers, etc.), and indicates that extrapolating the current injection rate to Gyr timescales is probably not a viable assumption. Taken together, these results directly support the multiple-step nature of the ICL build-up, with incremental but non-monotonous increases of its ICL fraction over its lifetime \citep{Rudick2006, rudick2009}. 

\subsection{Dynamical state}

ICL fractions in the visible/NIR bands can be used as a probe of cluster dynamical state, although their interpretation should be nuanced by the wavelength at which they are measured. ICL fraction colours with filters probing young stellar populations \citep[Hubble/ACS~F606W for example;][]{Jimenez-Teja2018, Dupke2022, Jimenez-Teja2023} are tightly connected to the cluster dynamical state, as recent dynamical mergers result in ICL fraction peaks in these wavelength regimes. This does not hold for the NISP filters, as they cover the spectral distribution of stars in the NIR, a wavelength regime insensitive to stellar age and rather depending on metallicity. One also expects new stars to be injected steadily into the ICL through the life of a cluster, resulting in an overall high broadband ICL fraction for well-evolved, relaxed galaxy clusters \citep{golden-marx2024}. The most extreme cases for this scenario are fossil systems \citep{Dupke2022}. 

The rather high ICL and ICL+BCG fractions (respectively $28\%$ and $34\%$ averaged over all filters) are therefore indicative of SMACS~J0723 being a well-evolved galaxy cluster that had time to build up its ICL. 
This is supported by the M$_{\rm 14}$ values, typical of evolved and BCG-dominated galaxy clusters, and is consistent with the BCG and X-ray centroid being completely aligned, with signs of cooling in the core of the intracluster medium \citep{Mahler2023}. 
This scenario is degenerated with another possibility, the recent injection of stars in the ICL boosting the ICL fractions \citep{Rudick2006}, supported by the presence of many tidal features and a very high current injection rate (see previous section). 
Considering that the inner ($r<100$\,kpc) ICL+BCG colours are consistent with a major merger between the BCG and the fourth brightest galaxy \citep{Montes2022jwst}, and that the X-rays feature some inhomogeneities outside the core, SMACS~J0723 indeed appears to have undergone recent dynamical activity, with a potential merger with another cluster or group along an axis close to the line of sight \citep{Mahler2023}.

However, a cooling core takes a few Gyr to form, a timescale much larger than the decay timescales of the tidal substructures seen in the ICL.
Additionally, the ICL fractions do not vary with wavelength, indicating very similar intracluster and galactic stellar populations when looking at the cluster as a whole. 
A major merger with another cluster and another ICL seems therefore unlikely, as stronger perturbations would be seen in the X-ray cluster core.
Another probable scenario here is therefore a recent (and still ongoing) internal series of major galaxy mergers between the BCG and the other brightest satellites -- a complex grinding phase with strong enough tidal consequences to produce the unique substructures seen in JWST's images. These are superimposed on an older, well-mixed ICL diffuse component, fed by previous stripping of the same galaxies experiencing the mergers.

\section{Conclusions}
\label{sec:conclusions}

In this work, the JWST/NIRCam deep images of the original release \citep{Pontoppidan2022} targetting the galaxy cluster SMACS~J0723 in six bands (F090W, F150W, F200W, F277W, F356W, and F444W) and processed for LSB science by \citet{Montes2022jwst} were used to study the ICL in the NIR. Additional care was given to clean up the short channel filters from instrumental scattering light (see Sect.~\ref{subsec:preprocessing}), and a full wavelet-based decomposition of the images in each band was performed using \texttt{DAWIS} \citep{Ellien2021}. The ICL, BCG and satellite galaxies were all extracted and modelled leveraging wavelet scale, source size and spatial position criteria, allowing to synthesize 2D maps for each component (see Sect.~\ref{sec:image_analysis}).

The ICL and ICL+BCG fractions were computed for each filter using the synthesized maps, displaying a flat trend with wavelength with an average of respectively 28\,\% and 34\,\% within a 400\,kpc radius. In the same way, the flux ratios between the BCG and the second to fourth members also display quite flat values with wavelength (M$_{\rm 12}\sim 1.2-1.3$, M$_{\rm 13}\sim1.3-1.6$ and M$_{\rm 14}\sim 2.3-2.5$ through all filters). These high ICL fractions and magnitude gaps indicate that SMACS~J0723 is an evolved galaxy cluster with a dominant BCG and a well-built-up and homogeneous ICL.

Five prominent substructures spanning the ICL were also studied: the W big loop, the NW bridge, the S tidal tail and the E and SE tidal streams. Taken all together, they make up for 10--12\% of the total ICL+BCG budget within a 400\,kpc radius. This is slightly higher than expected in average from simulations \citep{rudick2009}. Considering the cluster mass, the dynamical timescales of these substructures are a few Myr -- this yields an instantaneous injection rate of a few $\times 10^3 L_{\odot}.{\rm yr}^{-1}$, a value much higher than what is found in relaxed nearby clusters \citep{Mihos2017}.

Assembling all these pieces, SMACS~J0723 appears as a great example of the build-up of ICL through episodic growth phases through dynamical activity. However, the most probable scenario here resembles a series of galaxy major mergers between the BCG and the other brightest satellites rather than the injection of pre-processed ICL following a cluster-cluster merger.

\begin{acknowledgements}
Amaël Ellien acknowledges fundings by the CNES post-doctoral fellowship program.
This work was made possible by utilising the CANDIDE cluster at the Institut d’Astrophysique de Paris. The cluster was funded through grants from the PNCG, CNES, DIM-ACAV, the Euclid Consortium, and the Danish National Research Foundation Cosmic Dawn Center (DNRF140). It is maintained by Stephane Rouberol.
This research made use of Photutils, an Astropy package for detection and photometry of astronomical sources \citep{Bradley2024}. 
This work was partly done using GNU Astronomy Utilities (Gnuastro, ascl.net/1801.009) version 0.19. Work on Gnuastro has been funded by the Japanese Ministry of Education, Culture, Sports, Science, and Technology (MEXT) scholarship and its Grant-in-Aid for Scientific Research (21244012, 24253003), the European Research Council (ERC) advanced grant 339659-MUSICOS, the Spanish Ministry of Economy and Competitiveness (MINECO, grant number AYA2016-76219-P) and the NextGenerationEU grant through the Recovery and Resilience Facility project ICTS-MRR-2021-03-CEFCA.
\end{acknowledgements}

%
%

\bibliographystyle{aa} 
\bibliography{dawis_jwst}

\appendix

\section{Short channel instrumental light background}
\label{app:short_channel_scattered_light_background}

As stated by \citet{Montes2022jwst}, the short channel images were still significantly contaminated by instrumental instrumental light, despite their improved reduction pipeline. This was immediately confirmed by a first run of \texttt{DAWIS} on the images and by synthesizing ICL maps with simplistic selection schemes (see Section~\ref{subsec:atom_selection} and Appendix~\ref{app:atom_selection_schemes}): the ICL was deformed with linear features spatially corresponding with instrumental light stripes. Additional corrections were needed before applying \texttt{DAWIS} to remove this background. 

The following empirical approach was performed for each filter: first, a WS synthesized map with all atoms at wavelet planes lower than 7 is created and removed from the original image, leaving a residual image with large-scale variations. These large-scale variations are composed of leftover ICL and the instrumental light background. Here the instrumental light background is approximated by two components: a first component composed of thin stripes giving the sharp details visible in the original images, and a second inhomogeneous large-scale component with different values for each of the four corners corresponding to the four detectors. This second component is first estimated by masking pixels brighter than a $1\sigma$-median threshold (masking sharp edges, most of ICL and some sharp wavelet residuals) before computing an interpolated median background. This first background is subtracted from the residuals, which now contains only ICL leftover and instrumental light sharp edges. Mimicking the correction performed by \citet{Montes2022jwst}, a $3\sigma$-clipped median value is computed from each half-row, before performing the same on half-columns. This gives a sharp edge background for each corner of the image, which is then removed. The large-scale background is then reincorporated in the sharp-edge-free residuals, before masking bright pixels once again and performing a new estimation of the large-scale background. Finally, this last background is removed from the images, giving the input images for \texttt{DAWIS} to run again. 

The resulting combined background, alongside original, corrected and \texttt{DAWIS}-reconstructed images is shown in Figure~\ref{fig:plot_array_recim_short}. 
\rev{The improved background is not perfect as some stripe residuals are still visible, mainly in the F090W image. 
However, these residuals are mainly situated close to the image border and filtered out during the atom selection (see Sect.~\ref{app:atom_selection_schemes}).
The ICL linear deformations seen before instrumental light background correction are not present in the \texttt{DAWIS} outputs (see fourth column of Fig.~\ref{fig:plot_array_recim_short}) or in the synthesized ICL+BCG maps (see Fig/~\ref{fig:plot_example_recim_all_filter_maps}).}


\begin{figure*}
    \centering
    \includegraphics[width=\textwidth]{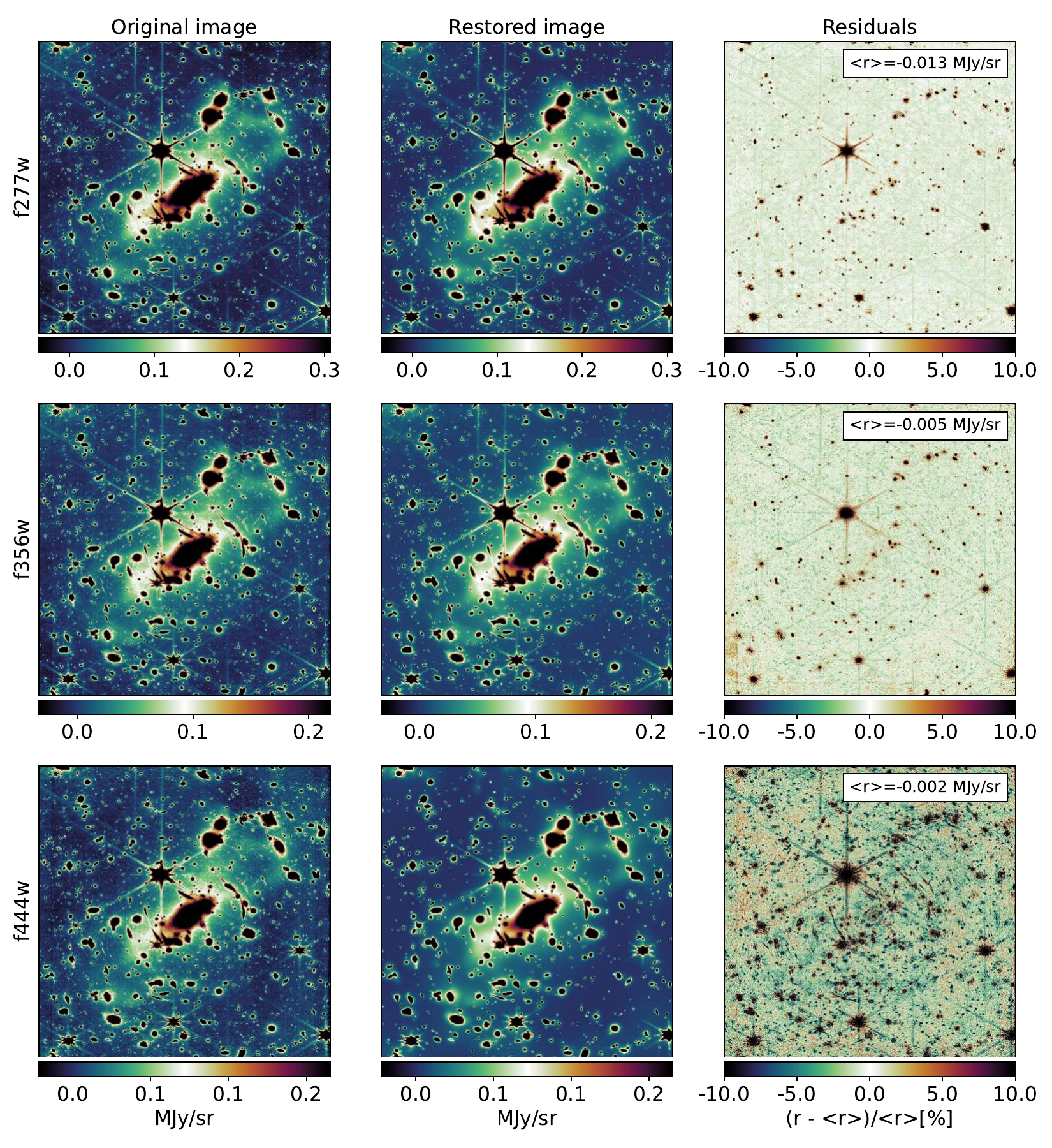}
    \caption{\rev{Same as Fig.~\ref{fig:plot_array_mix} but for the long channel filters F277W, F356W and F444W.}}
    \label{fig:plot_array_recim_long}
\end{figure*}
\begin{figure*}
    \centering
    \includegraphics[width=\textwidth]{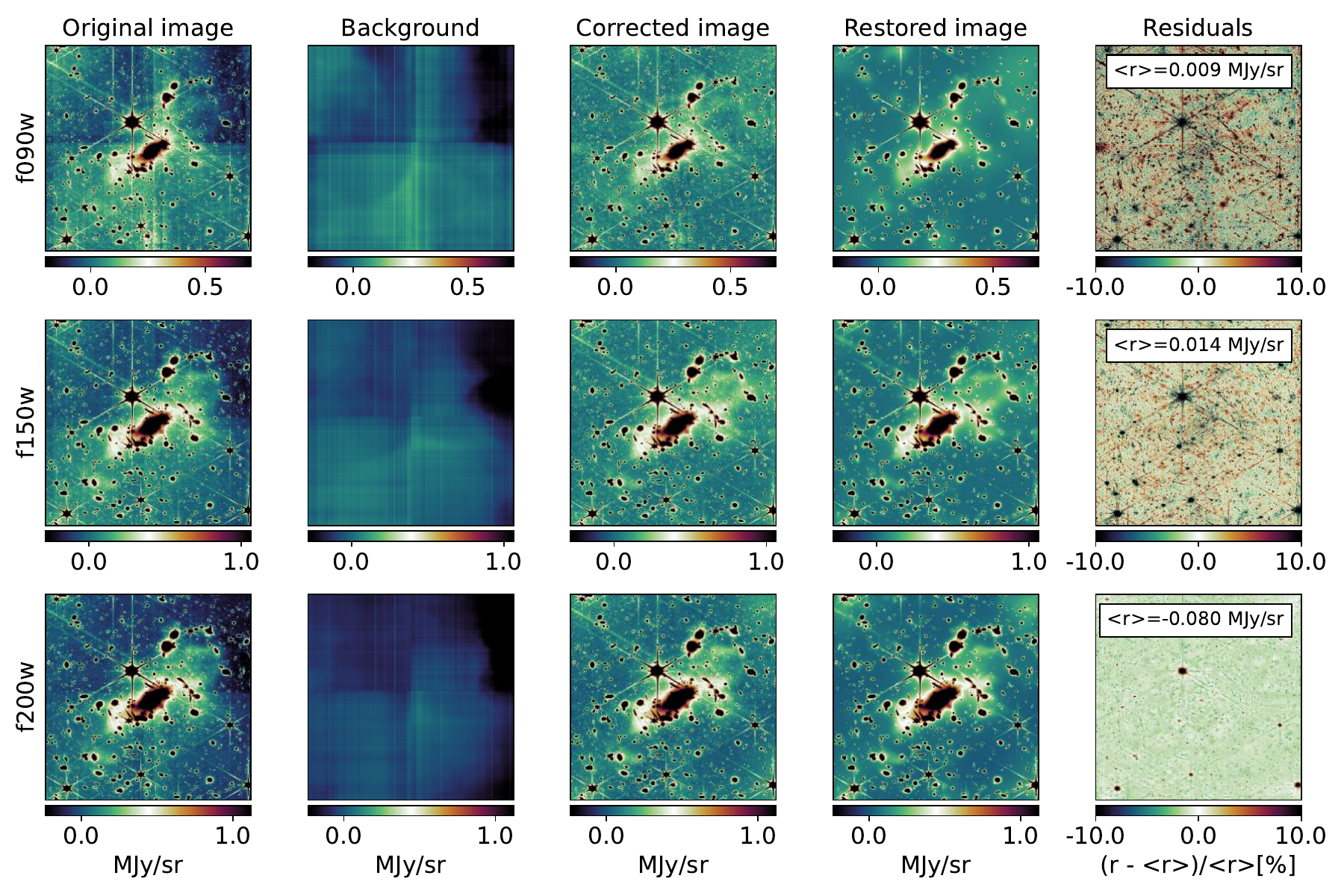}
    \caption{Outputs from \texttt{DAWIS} for all short channel filters. From left column to right column: original image, instrumental light background, background-subtracted image, restored image (synthesized by summing all atoms detected and restored by \texttt{DAWIS}), residuals (original image with restored image subtracted from).
    \rev{The size of the images is $\ang{;1.75;}\times\ang{;1.75;}$ ($\sim680\,\mathrm{kpc}\times680\,\mathrm{kpc}$ at the cluster's redshift).}
    The residuals are displayed with enhanced contrast (\rev{deviation of individual residual pixels $r$ from the median residual value $<r>$, expressed in percent)}.}
    \label{fig:plot_array_recim_short}
\end{figure*}

\section{DAWIS input parameters}
\label{app:dawis_input_parameters}

The operating mode of \texttt{DAWIS} is set for all images to standard values with a relative threshold $\tau=0.8$ and an attenuation factor $\gamma=0.5$, as advocated by E21. The maximum number of wavelet planes explored by the algorithm is set to 10 for long channel images, as their size is approximately $2^{11}$~pix $\times2^{11}$~pix. The convergence of the algorithm is controlled by a parameter $\epsilon=\frac{\sigma_{(i-1)} - \sigma_{(i)}}{n_{\mathrm{obj},i}\sigma_{(i-1)}}$, with $\sigma_{(i)}$ the standard deviation of the whole image and $n_{\mathrm{obj},i}$ the number of sources detected and restored at iteration $i$. Additional details about the form of this parameter are given in E21. However, note that in this work the average of $\epsilon$ over the last five iterations is taken for the convergence instead of the individual value for the last iteration. This prevents wrong convergence due to individual iterations where a great amount of flux is detected, restored and removed from the image. For this work, $\epsilon$ is set to $10^{-4}$ for all images.

\section{Atom selection schemes}
\label{app:atom_selection_schemes}


Three selection schemes are used in this work to separate atoms based on their properties:

i) Wavelet separation (WS): this scheme performs a separation of atoms based on the wavelet scale at which they are detected. Atoms detected at lower wavelet scales than a given hard limit $w_s$ are summed to synthesize a small detail image while atoms detected at higher ($w_s$ included) wavelet scales are summed to synthesize a large detail image. Since small sources (typically galaxies) are detected in the first wavelet scales before progressively leaving room to larger structures in lower frequency scales, this scheme allows a first order separation of ICL from smaller structures (E21).

ii) Spatial Filtering (SF): this scheme cross-correlates spatial information (galaxy coordinate catalog, masks...) with the coordinates of atom centroids to separate them into different components located in different areas of the image. This scheme is used to extract atoms associated to cluster galaxies for example, or again to remove atoms associated to foreground stars.

iii) Size separation (SS): this scheme performs a separation of atoms based on their physical size. Atoms with an x-axis and/or a y-axis size\footnote{The axes discussed here are the axes of the image.} smaller than a hard size threshold $S$ are kept as small objects while the others are kept as large objects.

\begin{figure*}
    \centering
    \includegraphics[width=\textwidth]{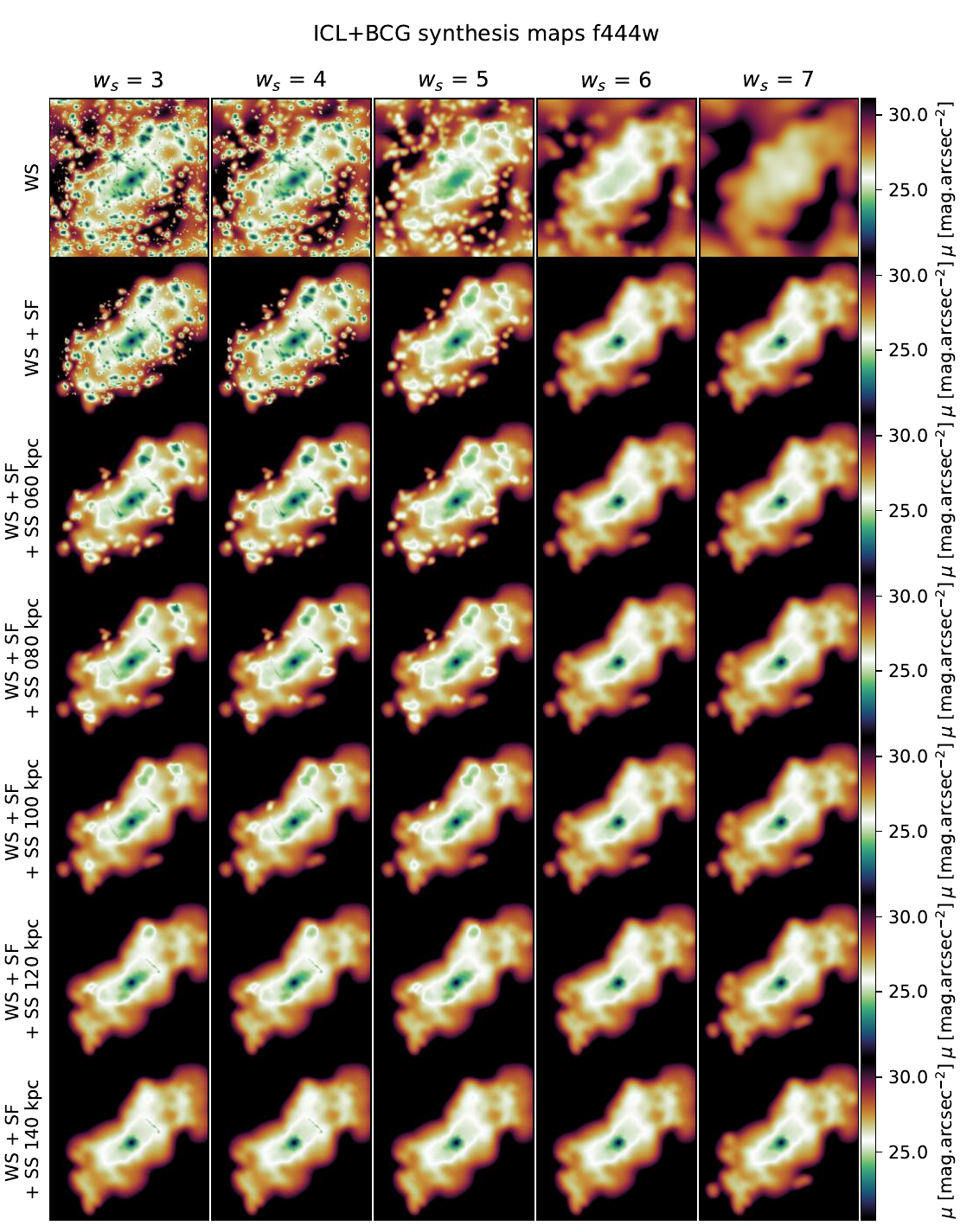}
    \caption{Grid of ICL+BCG synthesis models for the F444W filter. The vertical axis shows the name of the models, which are single or combinations of atom selection schemes: wavelet separation (WS), size separation (SS) and spatial filtering (SF). Different versions of the WS+SF+SS model are shown, with a characteristic size varying from 60 to 140\,kpc. The horizontal axis shows the different wavelet scales at which the WS is performed.}
    \label{fig:plot_array_bcgicl_maps_f444w}
\end{figure*}

The process of synthesizing images for this work is empirically performed by applying successive combinations of the WS, SF and SS schemes to atoms (the case of the F444W ICL+BCG synthesis image is shown in Figure~\ref{fig:plot_array_bcgicl_maps_f444w} as an example). First, a WS is applied with $w_s=\left[3, ..., 7\right]$, allowing a first rough separation of large sources from smaller features. The results are shown in the first line of Figure~\ref{fig:plot_array_bcgicl_maps_f444w}. An eye evaluation of the WS synthesis images allows to quickly see that most of galaxy and foreground starlight profiles are removed at $w_s=5$. However, the low-frequency signal is strongly contaminated by spurious structures such as galaxy/star extended light halos or again extended sources due to the borders of the image.

To remove these contributions a SF scheme is applied on top of the WS. A set of circular masks enclosing the core of foreground stars is created by hand, and atoms with centroid coordinates within these masks are removed completely from the analysis. In the same way, atoms with centroid coordinates outside a large handmade ellipse enclosing the ICL are also removed from the analysis. All atoms detected at wavelet scales lower than $w_s$ and with centroid coordinates within a radius of 10~kpc from cluster galaxy coordinates are selected and summed into a cluster galaxy (BCG included) synthesis image. In the case of the ICL+BCG image, the atoms centered on the BCG are simply removed from the galaxy image and added back to the ICL. The results for F444W are displayed on the second line of Fig.~\ref{fig:plot_array_bcgicl_maps_f444w}. 

On top of the WS+SF model, a SS is added to perform an additional separation of atoms based on their physical size. Contrary to the WS, limited by the amount of wavelet scales used for the analysis, a large number of size thresholds $S$ can be tested to remove/add atoms from the images. In this work a SS is applied to remove atoms with sizes lower than $S=\left[ 60, 80, 100, 120, 140 \right]$~kpc from the ICL and ICL+BCG synthesis images. The results can be seen on rows 3 to 7 in Figure~\ref{fig:plot_array_bcgicl_maps_f444w}.

Due to the fuzzy and ambiguous definition of ICL, one cannot determine exactly which of these models represents best the ICL. However, an eye estimation allows us to see that extreme models are either missing ICL signal (all models with $w_{\mathrm{s}}=7$) or displaying signal from galaxies (WS and WS+SF models with $w_{\mathrm{s}}=3$ or 4). By process of elimination, models such as the WS+SF+SS\,80\,kpc or WS+SF+SS\,100\,kpc with $w_{\mathrm{s}}=5$ or 6 are the best candidates to give the most accurate picture of the ICL. For simplicity, only the WS+SF+SS\,80\,kpc with $w_{\mathrm{s}}=5$ values are shown and discussed in the main text.

\section{Atom selection error}
\label{app:atom_selection_error}

While the synthesized models computed in the previous section are essential to visualize the ICL and get a general impression of its shape, they do not ensure an unbiased estimate of the ICL. The deterministic nature of the reconstruction process of \texttt{DAWIS} does not provide any error bar on the reconstructed atom properties and therefore provides no error bar on more global measurements such as ICL fractions. It does not allow the estimation of atom misclassification (non-ICL atoms included in ICL model, and missed ICL atoms) or the atom resolution of \texttt{DAWIS} (ICL and another source signals can be mixed within a single atom). This is true for the present work, where the high density of overlapping sources in the images leads to some spurious light contaminating the ICL or the satellite galaxy synthesis images (an example is light from the "Beret" galaxy contaminating an overlapping satellite galaxy). Some of these effects can also oppositely influence measurements, resulting in bias compensation (as already highlighted in E21).

To estimate the significance of these effects and to provide error bars, a randomized bootstrap method is applied to each measurement (fluxes, ICL fractions...): variants for each synthesized image are created by randomly adding/removing atoms. Both the number of atoms constituting the image and the atoms themselves can be modified. This allows to probe cases in which ICL is more/less contaminated by spurious sources and cases in which ICL is missing light. Formally, we consider $N$ atoms constituting a synthesized image resulting from a classification scheme and assume two parameters: $P_w$ the percent of 'wrongly classified' atoms and $P_l$ the percent error on the number of atoms composing the model. First, a new number $N_s$ of atoms is drawn uniformly within $\left[ N(1-P_l), N(1+P_l) \right]$. If $N_s\leq N$, the atoms composing the new model are drawn from the $N$ original atoms, probing cases where less flux is injected in the model respectively to the classification scheme (under-reconstructed cases). The cases with $N_s>N$ probe instances in which more flux is injected in the model respectively to the scheme (over-reconstructed cases). To emulate contamination, several wrongly classified atoms $N_w$ are drawn within $\left[ 0, N_s.P_w \right]$. Then $N_s-N_w$ atoms are drawn from the original $N$ atoms, while the $N_w$ contaminants are drawn from atoms that were not selected originally by the scheme (and can therefore include light from foreground/background galaxies or MW stars). This new atom list is then used to produce a variant synthesis image of the selection scheme.

This process is repeated $M$ times, resulting in a sample of $M$ models for each selection scheme. When measuring a quantity for a selection scheme, it is therefore taken as the median of the distribution while the upper and lower errors are computed as the 5th and 95th percentiles. For the rest of this work, $P_w=0.1$, $P_l=0.1$, and $M=100$.




\end{document}